\documentclass[onecollarge,natbib]{svjour2}
\bibpunct{[}{]}{;}{n}{}{,} 
\smartqed  
\usepackage{graphicx}
\usepackage{amsmath}

\journalname{Few-Body Systems (EFB22)}
\begin{document}

\title{Universality of three-body systems in 2D: parametrization of the bound states energies 
}


\author{F.~F. Bellotti \and
		T.~Frederico \and
		M.T.~Yamashita \and 
		D.V.~Fedorov \and
		A.S.~Jensen \and
		N.T.~Zinner
}


\institute{F.~F. Bellotti \at
              Instituto Tecnol\'{o}gico de Aeron\'autica, 12228-900, S\~ao Jos\'e dos Campos, SP, Brazil \\
              Department of Physics and Astronomy, Aarhus University, DK-8000 Aarhus C, Denmark \\
              Instituto de Fomento e Coordena\c{c}{\~a}o Industrial, 12228-901, S{\~a}o Jos{\'e} dos Campos, SP, Brazil \\
              Tel.: +55-12-39477147\\
              Fax: +55-12-39477111\\
              \email{ffbellotti@gmail.com}             \\
		\and
           T.~Frederico \at
              Instituto Tecnol\'{o}gico de Aeron\'autica, 12228-900, S\~ao Jos\'e dos Campos, SP, Brazil \\
        \and
           	M.T.~Yamashita  \at
    		  Instituto de F\'\i sica Te\'orica, UNESP - Univ Estadual Paulista, 01156-970, S\~ao Paulo, SP, Brazil \\
		\and
			D.V.~Fedorov, A.S.~Jensen, N.T.~Zinner \at
              Department of Physics and Astronomy, Aarhus University, DK-8000 Aarhus C, Denmark \\
}

\date{Received: date / Accepted: date}

\maketitle

\begin{abstract}
Universal properties of mass-imbalanced three-body systems in 2D are studied using zero-range interactions in momentum space. The dependence of the three-particle binding energy on the parameters (masses and two-body energies) is highly non-trivial even in the simplest case of two identical particles and a distinct one.  This dependence is parametrized for ground and excited states in terms of {\itshape supercircles} functions in the most general case of three distinguishable particles.

\end{abstract}

\section{Introduction}
\label{intro}
The behavior of three-boson systems changes remarkably from two (2D) to three dimensions (3D). The most famous example is the Efimov effect , which was predicted and measured for three identical bosons in 3D systems, but is absent in 2D. While in 3D the Efimov effect produces an infinite ladder of geometrically spaced states when the scattering length diverges, previous studies have shown that the spectrum of three identical bosons in 2D contains exactly one two-body and two three-body bound states in the limit where the range of the interaction goes to zero \cite{tjonPLB1975}.

Starting from the well-known case of three identical bosons, systematic calculation of universal properties of mass-imbalanced three-body systems in 2D are performed using zero-range interactions in momentum space. When at least one pairwise interaction is different from the other two, the dependence of the three-body energy on the two-body energy is quite different for ground and excited sates. For instance, whereas there is always a bound ground state for any three-body system in 2D (assuming at least two pairs are bound), the first excited state only occurs for special choices of masses and two-body binding energies. Nevertheless, three-body ground and excited state energies are parametrized as {\itshape supercircles} \cite{lame1818}, where the scaled two-body energies are the "coordinates" and the radius and the power represents a weak mass-dependence \cite{bellottiPRA2012}.  

\section{Formalism} 
\label{sec:1}
We investigate $ABC$ bound systems whose dynamics is restricted to two spatial dimensions (2D). The masses are $m_A,m_B,m_C$ and the pairwise interactions are described for attractive zero-range potentials. The three-body wave function $\left|\Psi_{ABC}\right\rangle$ for any $s-$wave bound state of energy $E_3$ is a solution of the free Schr\"oedinger equation, except where particles overlap. Using the Faddeev decomposition in momentum space, the bound state wave function is written as (we introduce here $(\alpha,\beta,\gamma)$ as cyclic permutations of $(A,B,C)$)
\begin{equation}
\left\langle \mathbf{q}_\alpha,\mathbf{p}_\alpha \right.\left|\Psi_{ABC}\right\rangle=\Psi\left(\mathbf{q}_\alpha,\mathbf{p}_\alpha\right)=\frac{f_{\alpha}\left(q_\alpha\right)+f_{\beta}\left(\left| \mathbf{p}_\alpha- \frac{m_\beta}{m_\beta+m_\gamma}\mathbf{q}_\alpha\right|  \right)+f_{\gamma}\left(\left| \mathbf{p}_\alpha+ \frac{m_\gamma}{m_\beta+m_\gamma}\mathbf{q}_\alpha\right| \right)}{|E_{3}|+\frac{q_\alpha^{2}}{2m_{\beta \gamma,\alpha}}+\frac{p_\alpha^{2}}{2m_{\beta \gamma}}} , 
\label{wave}
\end{equation}
where $\mathbf{q}_\alpha$ is the $\alpha$ particle momenta with respect to center-of-mass of the subsystem $\beta-\gamma$, $\mathbf{p}_\alpha$ is the pair relative momenta, $m_{\beta \gamma,\alpha}= m_\alpha(m_\beta+m_\gamma)/(m_\alpha+m_\beta+m_\gamma)$ and $m_{\beta \gamma}= (m_\beta+m_\gamma)/(m_\beta+m_\gamma)$ are the reduced masses and $f_{\alpha,\beta,\gamma}(\mathbf{q})$ are the Faddeev components, or spectator functions.

The three-body energy, $E_3$, is the solution of a set of three coupled homogeneous integral equations. This set of equations involves the three spectator functions and is written, in a compact form, as
\begin{equation}
f_{\alpha}\left( \mathbf{q}\right)  =\tau_\alpha(q,E_3) \int d^{2}k\left( \frac{f_{\beta}\left(k\right) }{|E_{3}|+\frac{q^{2}}{2 m_{\alpha \gamma}}+\frac{k^{2}}{2 m_{\beta \gamma}}+\frac{1}{m_\gamma}\mathbf{k}\cdot \mathbf{q}}+
\frac{f_{\gamma}\left( k\right) }{|E_{3}|+\frac{q^{2}}{2m_{\alpha \beta }}+\frac{k^{2}}{2m_{\beta \gamma}}+\frac{1
}{m_\beta}\mathbf{k}\cdot \mathbf{q}}\right) ,
\label{spec}
\end{equation}
where 
\begin{equation}
\tau^{-1}_\alpha(q,E_3)= 4\pi m_{\beta \gamma}\ln \left( \sqrt{\frac{\frac{q^{2}}{2m_{\beta \gamma,\alpha} }+|E_{3}|}{|E_{\beta\gamma}|}}\right)
\label{tau}
\end{equation}
are the matrix elements of the two-body T-matrix. The logarithmic term comes from the low-energy expansion of the scattering amplitude in 2D, as it was showed for equal mass particles in \cite{adhikariAJoP1986}. Besides, the logarithm leads to an ambiguity in the definition of the scattering length in 2D, which depends on the scale used to measure the energy. We use the binding energy of the pairs, $E_{\beta\gamma}$, as the  physical scales in the problem. 

Notice that three-body bound states are only allowed when at least two pairs are bound. For each non-interacting pair, the respective spectator function is zero. Whenever more than one pair is not interacting, the set of equations \eqref{spec} does not exist. This can be seen by setting $E_{\beta\gamma}=0$ in Eq.~\eqref{tau}, for the non-interacting $\beta-\gamma$ pair. In 2D, the effect of a non-interacting pair in Eq.~\eqref{spec} is the same as setting this pair with null interaction energy in Eq.~\eqref{tau}.

\section{Results and discussion}
\label{sec:2}
\begin{figure}[!h]
\centering
  \includegraphics[width=0.5\linewidth]{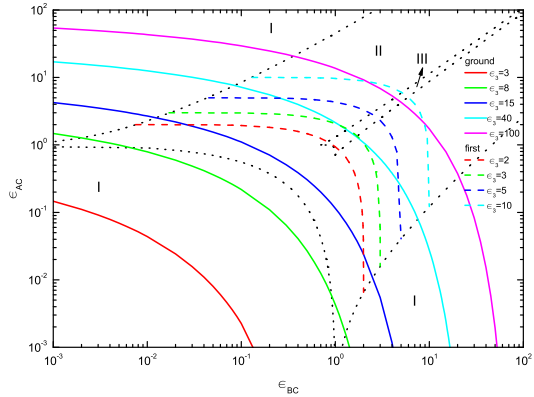}
  \caption{Contour diagrams with lines of fixed $\epsilon_3$ values as
  function of the two-body energies $\epsilon_{ac}$ and $\epsilon_{bc}$.
  The solid and dashed curves are for ground and excited states,
  respectively. Here $A$ is $^{87}$Rb, $B$ is $^{40}$K, and $C$ is 
  $^{6}$Li.
  The dotted curves are boundaries between regions where I, II, or III bound states  are allowed.
  }
\label{fig:1}       
\end{figure}

In the most general scenario of three distinguishable $ABC$ particles, the three-body binding energy ($E_3$) is a function of six independent parameters, namely masses and two-body binding energies. Realistic scenarios correspond to given particles with known masses, where pairwise interactions are tunable through Feshbach resonances. Common elements in cold atomic experiments are the alkali atoms. The particles are assumed to be $A=^{87}$Rb, $B=^{40}$K, and $C=^{6}$Li. Moreover, the large mass-asymmetry between $^{6}$Li and $^{87}$Rb allow the three-body system to have more bound states than a system composed of particles with comparable masses \cite{bellottiJoPB2013}. Once particles are chosen the number of independent parameters drops to three and defining the $AB$-binding energy ($E_{AB}$) as unit, the scaled three-body energy ($\epsilon_3=E_3/E_{AB}$) is a function of only two independent parameters, namely, the scaled $BC,AC$-energies. In other words, $\epsilon_3=F_n(\epsilon_{BC},\epsilon_{AC})$, where $n=0$ labels ground state and $n>0$ excited states. Full numerical calculation of  $\epsilon_3$ is shown in Fig.~\ref{fig:1} in the parametric space of the two-body scaled energies $\epsilon_{BC}$ and $\epsilon_{AC}$ for the chosen particles.

Although the equations which define the three-body energy as function of masses and two-body energies are very complicated, they can be expressed in a simple way, through the {\itshape supercircles} parametrization
\begin{equation}
\epsilon_{AC}^{t_{n}}+\epsilon_{BC}^{t_{n}}=R_{n}^{t_{n}} \; ,
\label{para}
\end{equation}
where the radius ($R_n$) and the exponent ($t_n$) are given by:	
\begin{equation}	
R_0(\epsilon_{3}) \approx 0.74 \epsilon_{3}-2.5\;\;,\;\; R_1(\epsilon_{3}) \approx \epsilon_{3}\;\;,\;\;t_0(\epsilon_{3}) \approx  \alpha_0 \frac{\epsilon_{3}^{p_0}+\beta_0}{\epsilon_{3}^{p_0} + \gamma_0},
\label{rt}
\end{equation}	
where $(p_0,\alpha_0) \simeq (0.04-0.06,0.3-0.5)$ exhibits a weak dependence on mass and $(\beta_0,\gamma_0) \simeq -(0.93-0.95),-(0.82-0.87)$ are almost mass independent and a similar expression can be calculated to $t_1(\epsilon_{3})$.

Notice that the lines describing $\epsilon_3$ in Fig.~\ref{fig:1}, which are the result of an extensive numerical work, are almost perfectly described by the parametrization in Eq.~\ref{para}. This result can be used for fast estimation of three-body energies and the number of bound states in specific systems for different sets of two-body energies.

As it can be seen in Fig.~\ref{fig:1}, the result presented in Eq.~\eqref{para} is quite general. It holds for any choice of two-body energies, including the case where one of them is equal zero. In this case, the properties of the system change \cite{bellottiJoPB2013}, but the parametrization is still valid. As it was discussed in Sec.~\ref{sec:1}, if more than one pair is noninteracting, there is no three-body bound state.

\begin{acknowledgements}
This work was partly supported by funds
provided by FAPESP (Funda\c c\~ao de Amparo \`a Pesquisa do Estado
de S\~ao Paulo) and CNPq (Conselho Nacional de Desenvolvimento
Cient\'\i fico e Tecnol\'ogico) of Brazil, and by the Danish 
Agency for Science, Technology, and Innovation.
\end{acknowledgements}


\begin{thebibliography}{30}
\bibitem[{{Tjon}(1975)}]{tjonPLB1975}
{Tjon} JA (1975) {Bound states of $^{4}$He with local interactions}. Phys.
  Lett. B 56:217--220

\bibitem[{Lam{\'e}(1818)}]{lame1818}
Lam{\'e} G (1818) Examen des diff{\'e}rentes m{\'e}thodes employ{\'e}es pour
  r{\'e}soudre les probl{\'e}mes de g{\'e}om{\'e}trie.

\bibitem[{{Bellotti} et~al(2012){Bellotti}, {Frederico}, {Yamashita},
  {Fedorov}, {Jensen}, and {Zinner}}]{bellottiPRA2012}
{Bellotti} FF, {Frederico} T, {Yamashita} MT, {Fedorov} DV, {Jensen} AS,
  {Zinner} NT (2012) {Supercircle description of universal three-body states in
  two dimensions}. Phys. Rev. A 85:025-601
  
\bibitem[{{Adhikari}(1986)}]{adhikariAJoP1986}
{Adhikari} SK (1986) {Quantum scattering in two dimensions}. Amer. J.
  Phys. 54:362
  
\bibitem[{{Bellotti} et~al(2013){Bellotti}, {Frederico}, {Yamashita},
  {Fedorov}, {Jensen}, and {Zinner}}]{bellottiJoPB2013}
{Bellotti} FF, {Frederico} T, {Yamashita} MT, {Fedorov} DV, {Jensen} AS,
  {Zinner} NT (2013) {Mass-imbalanced three-body systems in two dimensions}.
  J. Phys. B 46(5):055301

\end{thebibliography}

\end{document}